# A critical analysis of three near-infrared photometric methods of estimating distances to cataclysmic variables

Michael J. Gariety[a] and F. A. Ringwald[a]

[a] California State University, Fresno
Department of Physics
2345 E. San Ramon Ave., M/S MH37
Fresno, CA 93740-8031

**Abstract**

A critical analysis of three methods of estimating distances to cataclysmic variables (CVs) is performed. These methods, by Ak et al., Beuermann, and Knigge, all use near-infrared (*JHK* or $K_s$) magnitudes and the Barnes-Evans relation. We compare all three methods to distances measured by trigonometric parallax by Thorstensen, with *Hubble Space Telescope*, and with the *HIPPARCOS* spacecraft.

We find that the method of Ak et al. works best overall for all CVs, predicting distances on the average 4% less than those measured by trigonometric parallaxes. The method of Beuermann overestimates distances by 52%. The method of Knigge underestimates distances by 26%, although this was only ever meant as a lower limit, since it assumes all light comes from the secondary star.

## 1.1 Introduction

Cataclysmic variables (CVs) are interacting binary star systems with orbital periods between 78 minutes and around 15 hours. For reviews, see Warner (1995) and Hellier (2001). In all CVs, the primary star is a white dwarf. The secondary star, often referred to as the donor star, is usually a K-M star that is filling its Roche lobe and transferring gas onto the primary star. While the secondary star is usually approximately on the main sequence, some are significantly evolved. The in-falling gas forms in most cases an accretion disk around the white dwarf. If the white dwarf exhibits any magnetism, the accretion disk may be altered by the magnetic field lines. A CV's luminosity arises from a combination of light from the primary, the secondary, and the accretion disk.

A CV's mass-losing secondary star determines the path of the CV's evolution. The size of the Roche lobe and therefore the mass and radius of the mass-losing secondary star correlates with the CV's orbital period. Warner (1995) and Hellier (2001) both discuss this and explain the correlation.

CVs are thought to evolve from long to short orbital periods, as they lose angular momentum over time. In the distribution of orbital periods, there is a long-period cutoff, a period gap, a period minimum, and period bouncers: see Patterson (1998) and Hellier (2001). Since there is a mass limit on the white dwarf, the Chandrasekhar limit of ~1.4 solar masses, and also because relatively few CVs have evolved secondary stars, few CVs have orbital periods greater than 12 hours.

The Period Gap is an abrupt drop in the number of systems in the range of 2-3 hours. For orbital periods of < 2 hours, CV mass transfer is thought to be driven primarily by angular momentum loss due to gravitational radiation. For orbital periods of > 3 hours, CV mass transfer is thought to be driven by magnetic braking. The mass transfer in CVs is thought to switch off as they evolve through the gap, because of changes in the secondary stars's magnetic field (Rappaport et al., 1983; Knigge et al., 2011).

There is also a sudden cutoff at a minimum of about 78 minutes, named the Period Minimum. This is due to the secondary star becoming degenerate, so that it is not supported by gas pressure, but by the quantum-mechanical requirement that adjacent atoms cannot be too close (see Hellier, 2001). At this point, the CV evolves to longer orbital periods. CVs that have evolved through the Period Minimum are called Period Bouncers (see Patterson, 1998).

**1.2 Distance Measurement by Trigonometric Parallax**

Distances for CVs are of fundamental importance, and are most reliably measured by trigonometric parallax. The Fine Guidance Sensors on *Hubble Space Telescope* (*HST*) allow for precise measurements of stellar parallax. While *HST* parallaxes are precise, *HST* observing time is limited and relatively few objects can be observed. Still, ten CVs have had trigonometric parallaxes measured with *HST* (Harrison et al., 1999; Harrison et al., 2004; McArthur et al., 1999; McArthur et al., 2001; Beuermann et al., 2003; Beuermann et al., 2004). Trigonometric parallaxes of three other CVs, IX Vel, V3885 Sgr, and AE Aqr, were measured by the *High Precision Parallax Collecting Satellite* (*HIPPARCOS*) (Duerbeck, 1999). *HIPPARCOS* also observed V603 Aql, QU Car, RR Pic, and RW Sex, but we have excluded their *HIPPARCOS* parallaxes from this study due to their low quality.

Professor John Thorstensen, of Dartmouth College, has produced a sample of ground-based CCD parallaxes for 26 CVs (Thorstensen 2003; Thorstensen et al. 2008). Although he admits estimations and small errors, we use these and the *HST* and *HIPPARCOS* distance measurements as a standard by which other CV distance estimation methods are compared. Three of these 26 CVs, WZ Sge, YZ Cnc, and SS Aur, also have trigonometric parallaxes measured by *HST* (Harrison et al. 2004). We have taken the *HST* parallaxes to supersede those by Thorstensen (2003), although as he notes, his agree with those from *HST*. We have excluded other ground-based trigonometric parallaxes, including those for AE Aqr, V603 Aql, RR Pic, and RW Sex (van Altena et al., 1995, also listed by Duerbeck, 1999), and for HX Peg (Dahn et al., 1988), because all are either superseded by *HST* measurements or have errors comparable to or larger than the parallaxes.

## 1.3 Distance Estimation With The Barnes-Evans Relation

We will use the trigonometric parallaxes described above as a standard for evaluating three other, near-infrared, methods of estimating distances to CVs: the methods of Ak et al (2007), Beuermann (2006), and Knigge (2006, 2007). The Barnes-Evans relation is fundamental to all three methods. It states that, for late-type stars on the main sequence, the surface brightness in the near-infrared is nearly constant for a wide range of spectral types. The near-infrared magnitude, especially in the $K$ band, correlates with the star's absolute magnitude and its radius (see Barnes and Evans, 1976).

Bailey (1981) first used the Barnes-Evans relation to estimate distances to CVs. His method assumes 100% of the light comes from the secondary star, thus giving only a lower limit to the distance to a CV. The CV's accretion disk can also contribute significantly to the $K$-band light. Berriman et al. (1985) discussed this, and presented a larger dataset of infrared photometry of CVs. Sproats et al. (1996) tried to improve on it, by considering how different fractions of light from the disk and the secondary star affect the infrared colors of the entire system.

## 1.4 The Distance Modulus and Interstellar Absorption

The distance modulus formula relating apparent magnitude, absolute magnitude, distance, and interstellar absorption is:

$$V - M_v = 5 \log (d) - 5 + A_v \qquad (1)$$

where:

$V$ = the apparent magnitude in the visual band,

$M_v$ = the absolute magnitude in the visual band,

$d$ = the distance in parsecs,

$A_V = 3.1\ E\ (B - V)$ = the interstellar absorption in the visual band, $\qquad (2)$

and:

$E(B-V)$ is the color excess (see page 382 of Cox, 2000).

We have listed in Table 1 all values for color excess used in this paper, all of which we took from estimates of Bruch and Engel (1994), Ringwald et al. (1996), and Harrison et al. (2004). Where no values for color excess or interstellar absorption are available, we assumed a zero value.

## 2.1 Methods

**Ak et al. (2007)**

The underlying principle of the method of Ak et al. (2007) relates the absolute magnitude in any wavelength interval with the orbital period and at least one color of the system. They called this the period-luminosity-color (PLC) relation. The luminosities and the colors used are those empirically measured for the entire CVs, not just their secondary stars. Ak et al. (2007) used $J$, $H$, and $K_s$ magnitudes from the Point-Source Catalog and Atlas (Cutri et al., 2003), which is based on the 2MASS (Two Micron All Sky Survey) observations. One advantage of using these near-infrared magnitudes and colors is that much of the light in these photometric bands comes from the secondary star; another is that the effect of interstellar reddening is weaker than that in visual wavelengths.

Ak et al. (2007) gave an absolute magnitude calibration and the following empirical formula:

$$M_J = -0.894 + -5.721 \log P_{orb} \text{ (day)} + 2.598 (J-H)_0 + 7.380 (H-K_s)_0 \qquad (3)$$

The distance modulus formula (1) was then used to determine distances for CVs. Figure 1 shows the distances estimated by Ak et al. (2007) for the CVs with measured trigonometric parallaxes. Figure 2 shows accompanying distance residuals, found by subtracting the distance estimates of Ak et al. (2007)

from the trigonometric parallaxes.

Ak et al. (2007) estimated the interstellar absorption for *J*, *H*, $K_s$, and *V* bands by using the equations of Fiorucci and Munari (2003):

$$A_J = 0.887\, E(B-V) \tag{4}$$

$$A_H = 0.565\, E(B-V) \tag{5}$$

$$A_{K_s} = 0.382\, E(B-V). \tag{6}$$

**Beuermann (2006)**

Beuermann begins with an empirical calibration of the surface brightness of stars as a function of color, building on the work of Barnes and Evans (1976), Bailey (1981), and Ramseyer (1994). This calibration allows a determination of the radius of a star of known distance and vice versa.

A potential source of significant error here is that this method assumes simultaneous measurement of *V* and *K*. Few instruments can do this, however, and CVs can change magnitude and color radically within hours or even minutes: see, for example, Berriman et al. (1985).

Beuermann used a third degree polynomial fit for the surface brightness:

$$S_K = 2.739 + 0.39318\,(V-K)^1 - 0.02433\,(V-K)^2 + 0.00150594\,(V-K)^3 \tag{7}$$

Distances were obtained from:

$$\log d = [(m_K - A_K - S_K)/5] + 1 + \log(R_{\text{eff}}/R_{\text{Sun}}) \tag{8}$$

where $m_K$ = observed apparent magnitude in the *K* band.

Beuermann used the equation by Patterson (1984) for relating $R_{eff}/R_{Sun}$:

$$R_{eff}/R_{Sun} = R/R_{Sun} = 0.427[P(hr)/4]^{1.073} \qquad (9)$$

where $P(hr)$ = orbital period, in hours.

Beuermann's method for obtaining distances to CVs uses the Caltech (CIT) photometric system for $J$, $H$, and $K$ magnitudes. In order to use 2MASS photometry, we used the following color transformations by Carpenter (2001) to convert between the 2MASS system to the CIT system:

$$(K_S)_{2MASS} = K_{CIT} + (0.000 \pm 0.005)(J - K)_{CIT} + (-0.024 \pm 0.003), \qquad (10)$$

$$(J - H)_{2MASS} = (1.076 \pm 0.010)(J - H)_{CIT} + (-0.043 \pm 0.006), \qquad (11)$$

$$(J - K_S)_{2MASS} = (1.056 \pm 0.006)(J - K)_{CIT} + (-0.013 \pm 0.005), \qquad (12)$$

$$(H - K_S)_{2MASS} = (1.026 \pm 0.020)(H - K)_{CIT} + (0.028 \pm 0.005). \qquad (13)$$

Figure 3 shows the distances we have estimated using the method of Beuermann (2006) for CVs with known trigonometric parallaxes, and Figure 4 shows accompanying distance residuals.

**Knigge (2006, 2007)**

Knigge's method tests how well distances may be estimated just from infrared magnitudes that are calculated from theoretical models of the secondary stars and orbital periods, together with the assumption that the secondary stars fill their Roche lobes. Like Beuermann, Knigge uses the CIT system, so we transformed his infrared magnitudes into the 2MASS system with the equations of Carpenter (2001). Knigge constructs a complete, semi-empirical donor sequence for CVs that is based on an understanding of donor properties and is consistent with all key observational constraints. By design, the donor sequence also reproduces the observed locations of the period gap, the period minimum, and the period bouncers. Distances are estimated using the distance modulus equation (1), in the *K*-band. Figure 5 shows the distance limits we have estimated with the method of Knigge (2006) for CVs with known trigonometric parallaxes, and Figure 6 shows accompanying distance residuals.

## 3.1 Results

Of the 26 CVs in Thorstensen's distance estimate studies, only two lacked 2MASS *JHK$_s$* apparent magnitudes, V396 Hya and HV Vir. We plotted the results of the methods of Ak et al. (2007), Beuermann (2006), and Knigge (2006, 2007). Figures 1, 3, and 5 respectively show the results of comparing the distance estimates of these three near-infrared photometric methods to trigonometric parallaxes by Thorstensen (2003, 2008), *HST*, and *HIPPARCOS*. Figure 7 shows all the data together on the same plot. Table 1 lists results for all near-infrared distance estimation methods with associated errors. Table 2 lists average residuals and percent differences between parallax distances and the methods of Ak et al. (2007), Beuermann (2006), and Knigge (2006, 2007). In order to compute reliable values of residuals and percent differences, GP Com, AH Her, V396 Hya, and HV Vir were omitted. GP Com is not really a CV, it is an AM CVn star, with a white dwarf being fed by another white dwarf (Ritter and Kolb 2003). AH Her has such large parallax errors that its parallax distance may be incorrect. V396 Hya is also a double degenerate, like GP Com (Ritter and Kolb 2003), and it and HV Vir lack *JHK$_s$* measurements from the 2MASS survey (Cutri et al. 2003).

**Conclusions**

The method of Ak et al. (2007), when compared to trigonometric parallax, works surprisingly well. From the percent differences of distances, the overall average is 4% less than the parallax distance. This underestimation is larger below the period gap, giving an average percent difference of 18% less than the parallax distance, rather than above the period gap, where distances are 11% above the parallax distances.

The method of Beuermann (2006) overestimates the distances with an average of 52% longer than that of parallax distance. For orbital periods above and below the period gap, the estimations of Beuermann (2006) are higher by 18% and 84%, respectively.

The method of Knigge (2006, 2007) also gives a lower limit for distance, because it assumes that all *JHK* light comes from the secondary star. The overall average for distance by Knigge (2006, 2007) is 26% less than that of parallax distance. Below the period gap, his method underestimates the distance by 36%; above the gap, his method underestimates the distance by 13%. This method makes no distance estimate for the GP Com due to its having an orbital period less than the lower limit of the semi-empirical donor sequence, because GP Com is an AM CVn star. Also, SS Cyg, Z Cam, RU Peg, and AE Aqr are omitted, due to their orbital periods being longer than the upper limit of the semi-empirical donor sequence. It may be possible to improve distance estimates from this method if there were a reliable way to estimate the fraction of *K*-band light that comes from the secondary stars, and the fraction that comes from other parts of the CV, particularly the disk.

**Acknowledgments**

This publication makes use of data from the Two Micron All Sky Survey, a joint project of the University of Massachusetts and the Infrared Processing and Analysis Center/California Institute of Technology, funded by the National Aeronautics and Space Agency and the National Science Foundation. This research made use of the SIMBAD database, operated at CDS, Strasbourg, France. This research also made use of the NASA Astrophysics Data System (ADS).

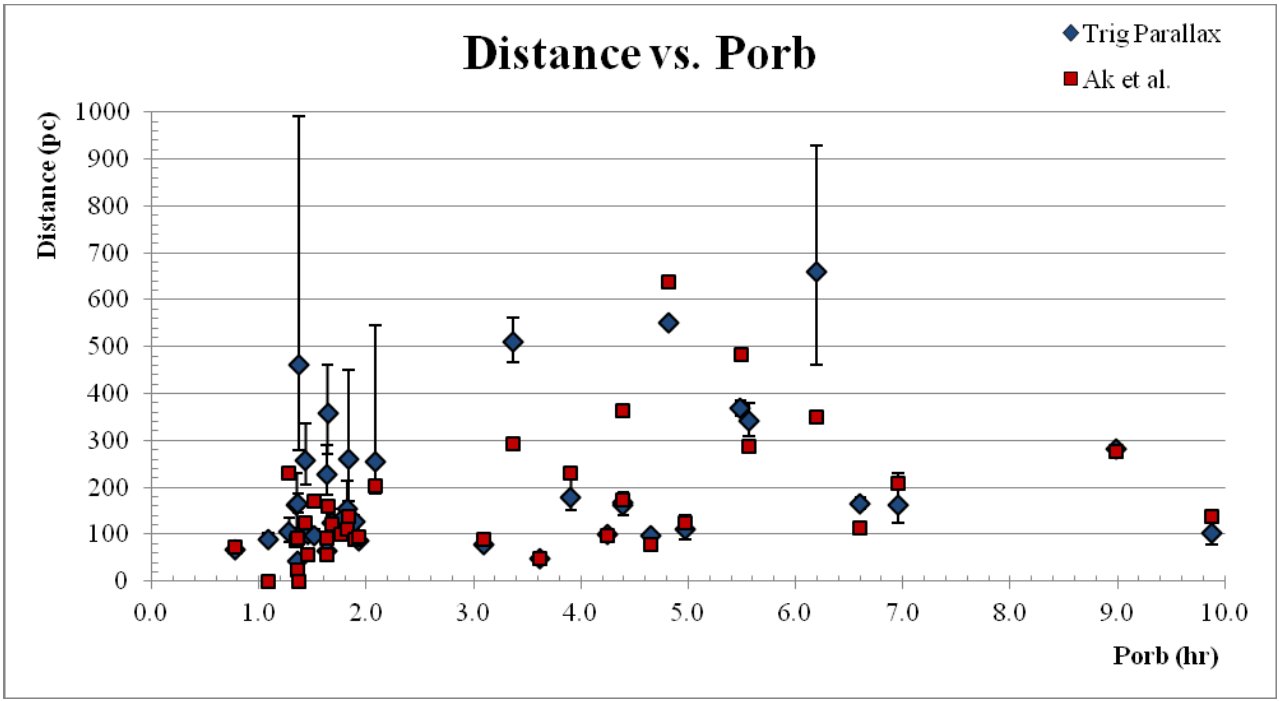

Figure 1: Distances to CVs as a function of orbital period, comparing trigonometric parallax with Ak et al. (2007). Trigonometric parallax data are shown with associated error bars. The largest errors belong to HV Vir at 1.37 hr and AH Her at 6.19 hr.

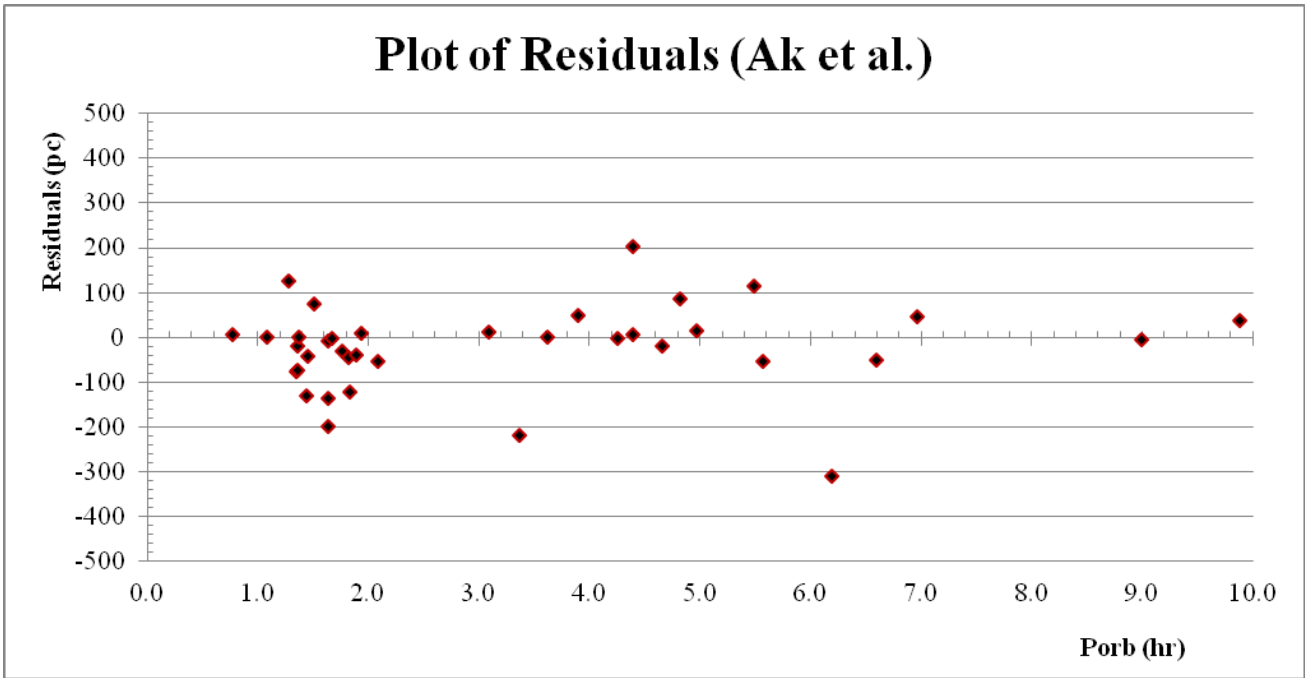

Figure 2: Residuals between the distance estimates of Ak et al. (2007) and trigonometric parallax.

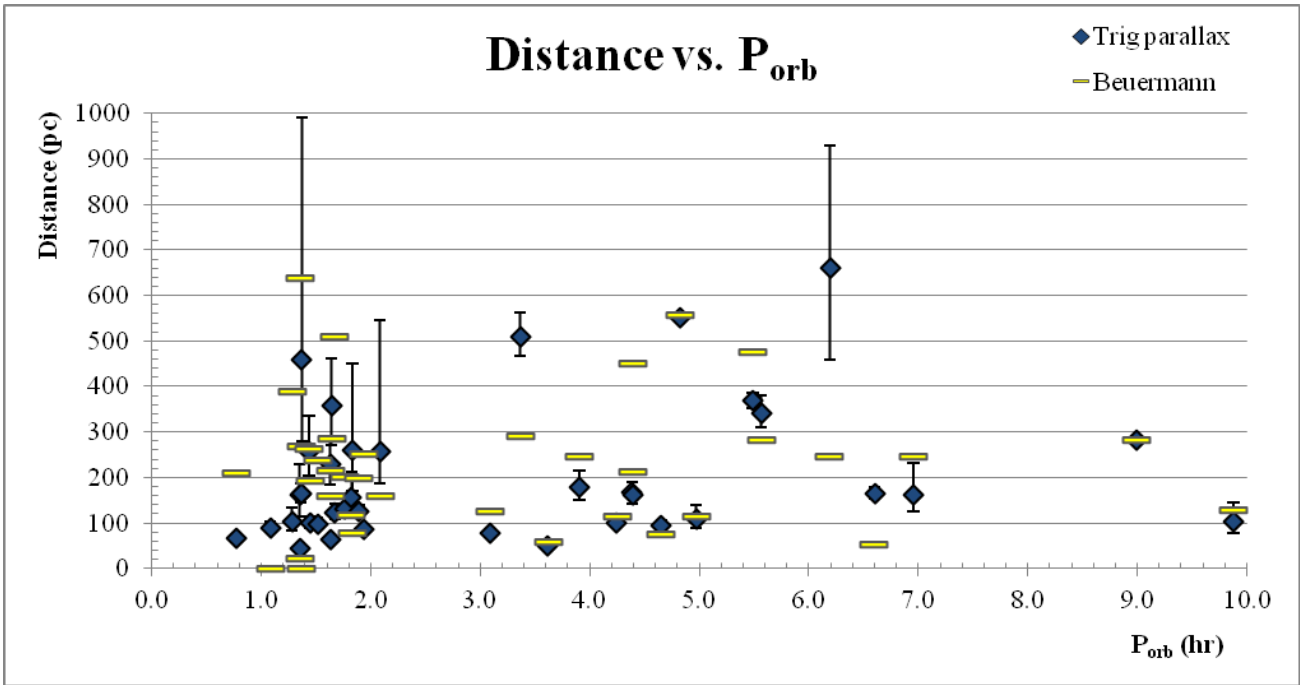

Figure 3: Distances to CVs as a function of orbital period, comparing trigonometric parallax with Beuermann (2006). Trigonometric parallax data are shown with associated error bars.

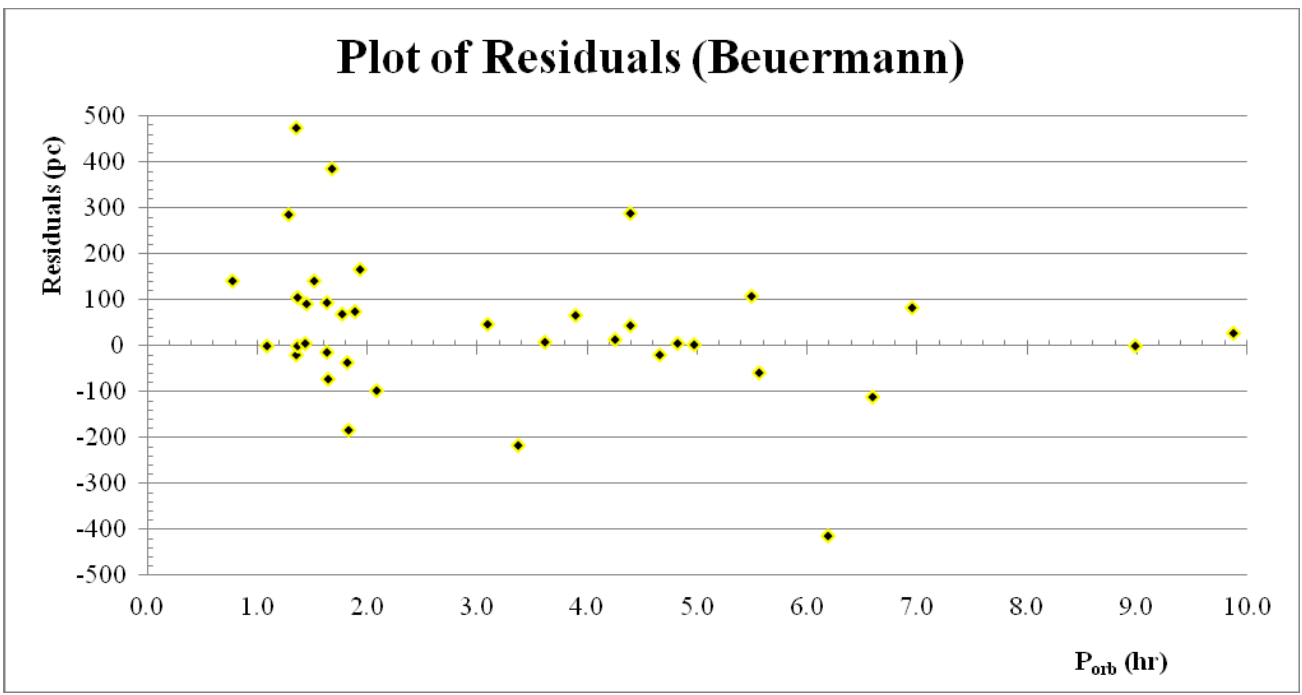

Figure 4: Residuals between the distance estimates of Beuermann (2006) and trigonometric parallax.

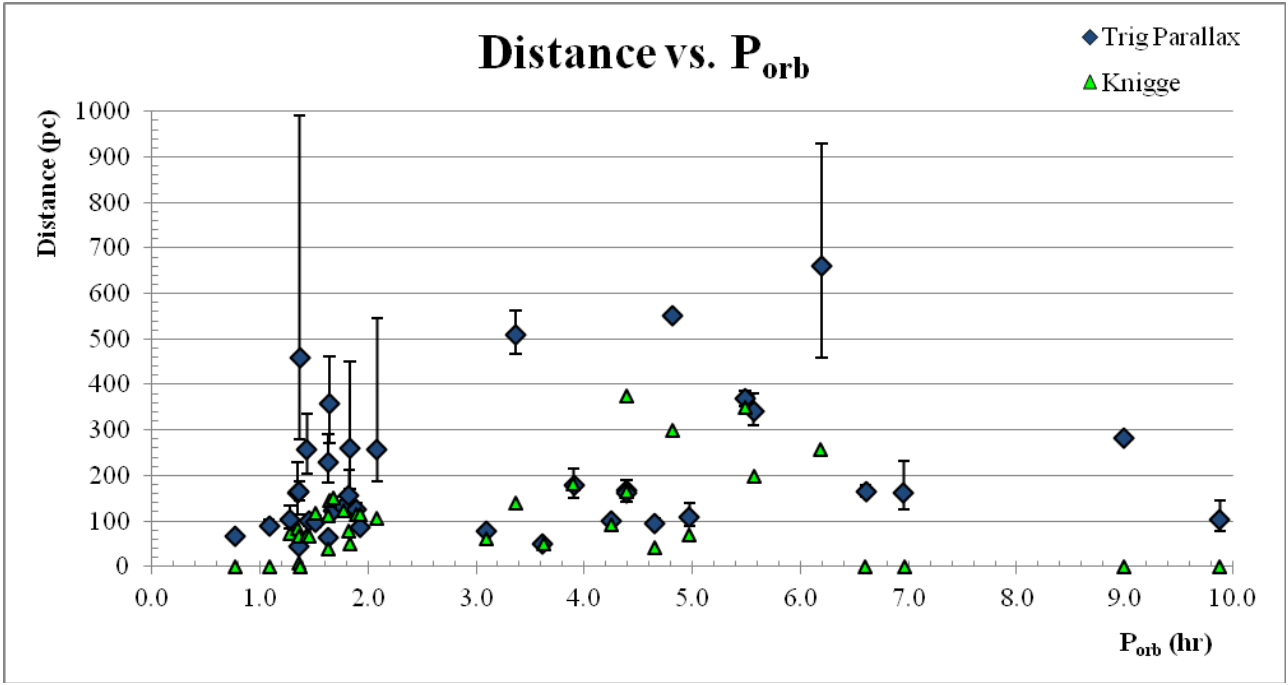

Figure 5: Distances to CVs as a function of orbital period, comparing trigonometric parallax with Knigge (2006, 2007). Trigonometric parallax data are shown with associated error bars.

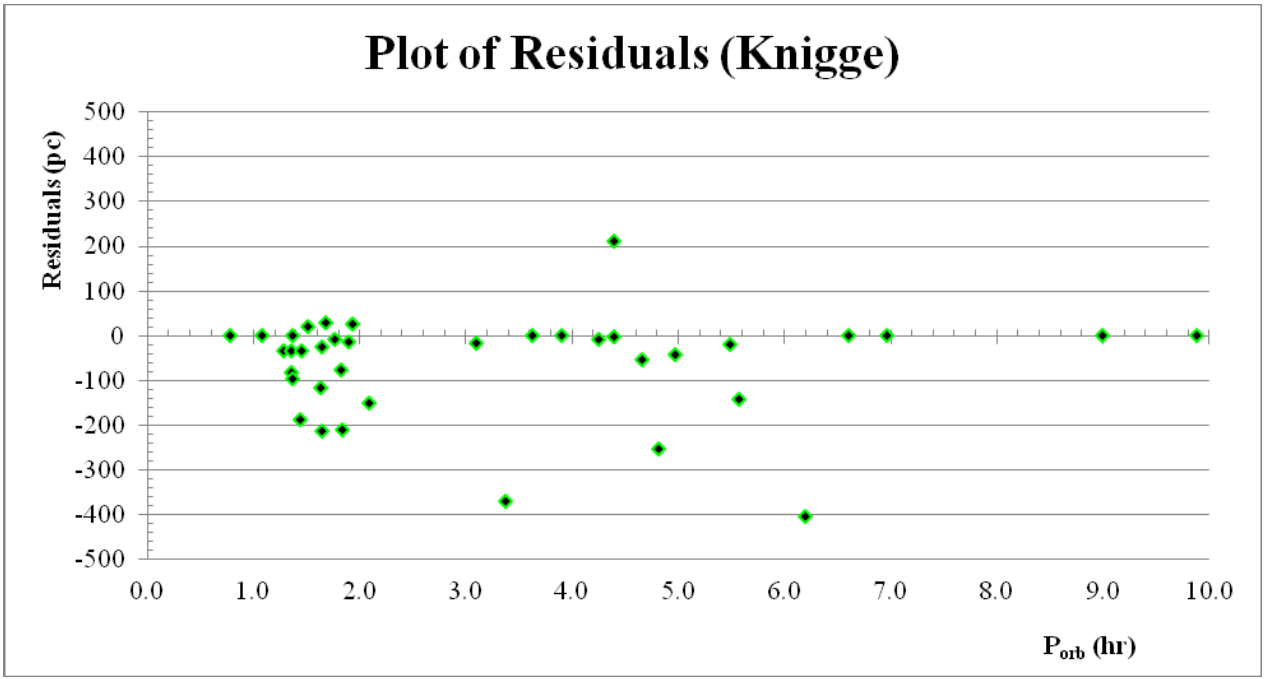

Figure 6: Residuals between the distance estimates of Knigge (2006, 2007) and trigonometric parallax.

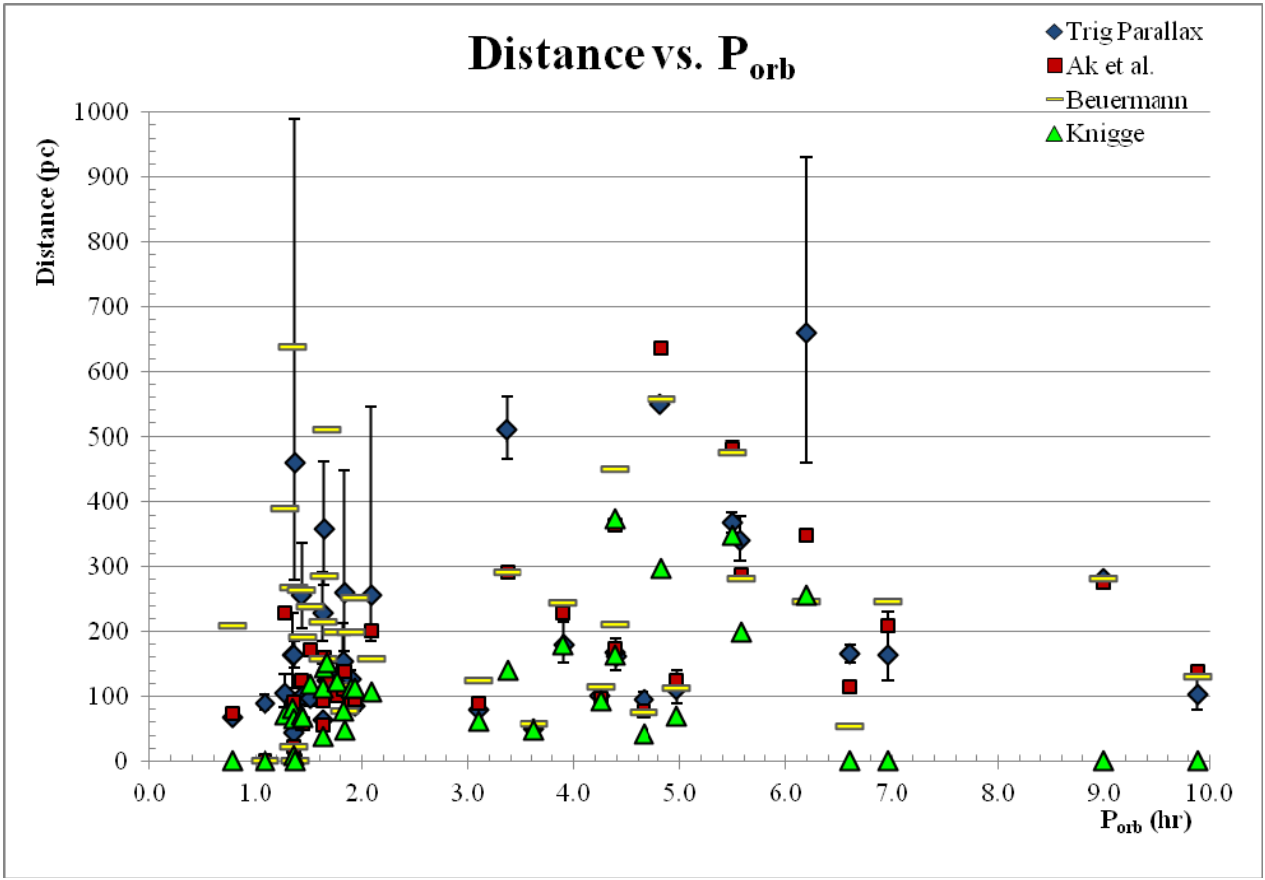

Figure 7: Distances to CVs as a function of orbital period. A comparison of the methods of trigonometric parallax with Ak et al. (2007), Beuermann (2006), and Knigge (2006, 2007). Distances from trigonometric parallaxes are shown with associated error bars.

| Name | $P_{orb}$ (days) | Parallax d (pc) | + | - | Parallax ref. | Ak d (pc) | Beuermann d (pc) | Knigge d (pc) | $E(B-V)$ | $E(B-V)$ ref. |
|---|---|---|---|---|---|---|---|---|---|---|
| **GP Com** | 0.032339 | 68 | 7 | 6 | a | 73 | 208.6 | – | – | |
| **V396 Hya** | 0.04519 | 90 | 12 | 10 | b | – | – | – | – | |
| **GW Lib** | 0.05332 | 104 | 30 | 20 | a | 229.7 | 389.7 | 71 | 0.00 | j |
| **EF Eri** | 0.056266 | 163 | 66 | 50 | a | 85.1 | 636.9 | 82.2 | – | |
| **WZ Sge** | 0.056688 | 43.5 | 0.3 | 0.3 | d | 22.8 | 23 | 8.3 | 0.00 | d |
| **SW UMa** | 0.056815 | 164 | 22 | 19 | b | 91.6 | 268.6 | 66.2 | 0.00 | k |
| **HV Vir** | 0.057069 | 460 | 530 | 180 | a | – | – | – | – | |
| **DW Cnc** | 0.059793 | 257 | 79 | 52 | b | 125.1 | 263.5 | 70.7 | – | |
| **T Leo** | 0.05882 | 101 | 13 | 11 | a | 57.6 | 191.3 | 67.9 | 0.00 | k |
| **VY Aqr** | 0.06309 | 97 | 15 | 12 | a | 171.8 | 238.1 | 118.1 | 0.07 | k |
| **BZ UMa** | 0.06799 | 228 | 63 | 43 | b | 92.5 | 214.9 | 112.2 | – | |
| **EX Hya** | 0.068234 | 64.5 | 1.2 | 1.2 | e | 56.1 | 158.5 | 39 | 0.00 | k |
| **IR Gem** | 0.0684 | 358 | 104 | 86 | b | 160.3 | 285 | 144.9 | 0.00 | k |
| **VV Pup** | 0.069747 | 124 | 17 | 14 | b | 122.7 | 510.2 | 151.9 | – | |
| **HT Cas** | 0.073647 | 131 | 22 | 17 | b | 101.1 | 199.7 | 123.4 | 0.03 | k |
| **V893 Sco** | 0.075961 | 155 | 58 | 34 | a | 111.1 | 117.3 | 77.1 | – | |
| **SU UMa** | 0.07635 | 260 | 190 | 90 | a | 138.7 | 77 | 48.8 | 0.00 | k |
| **MR Ser** | 0.078798 | 126 | 14 | 12 | b | 87.9 | 199.4 | 113.1 | – | |
| **AR UMa** | 0.080501 | 86 | 9 | 8 | b | 95 | 251.6 | 113.2 | – | |
| **YZ Cnc** | 0.0868 | 299 | 47 | 35 | d | 202.1 | 158.4 | 106.3 | 0.00 | k |
| **AM Her** | 0.128927 | 79 | 8 | 6 | a | 89.7 | 124.7 | 61.8 | – | |
| **V1223 Sgr** | 0.140244 | 510 | 52 | 43 | f | 292.1 | 291.7 | 140.2 | 0.15 | k |
| **QS Vir** | 0.150758 | 49 | 4 | 4 | b | 48.9 | 58.3 | 48.7 | – | |
| **KT Per** | 0.162658 | 180 | 36 | 28 | b | 229.5 | 245.4 | 180.2 | 0.18 | k |
| **U Gem** | 0.176906 | 100 | 4 | 4 | d | 96.7 | 114.4 | 92.6 | 0.04 | k |
| **SS Aur** | 0.1828 | 167 | 10 | 9 | d | 174.3 | 211.7 | 163.8 | 0.08 | k |
| **MQ Dra** | 0.18297 | 162 | 27 | 21 | b | 363.8 | 451.6 | 374.8 | – | |
| **IX Vel** | 0.193927 | 96 | 10 | 8 | i | 77.9 | 74.9 | 42.5 | 0.01 | k |
| **V3885 Sgr** | 0.207161 | 110 | 30 | 20 | i | 124 | 113.5 | 68.8 | 0.02 | k |
| **TV Col** | 0.228600 | 368 | 17 | 15 | h | 482 | 475.8 | 349.5 | 0.02 | k |
| **RW Tri** | 0.231883 | 341 | 38 | 31 | g | 287.5 | 282.2 | 199.3 | 0.10 | k |
| **AH Her** | 0.258116 | 660 | 270 | 200 | a | 349.2 | 247.2 | 255.7 | 0.03 | k |
| **SS Cyg** | 0.275130 | 166 | 13 | 13 | c | 114.1 | 53.9 | – | 0.07 | k |
| **Z Cam** | 0.289841 | 163 | 68 | 38 | a | 209.8 | 246.9 | – | 0.02 | k |
| **RU Peg** | 0.3746 | 282 | 0 | 0 | d | 275.5 | 282.1 | – | 0.00 | k |
| **AE Aqr** | 0.411656 | 102 | 42 | 23 | i | 138.8 | 129.5 | – | 0.00 | k |

Table 1: Table of CVs, their orbital periods, distances by trigonometric parallax with associated errors and references, distance estimates by Ak et al. (2007), Beuermann (2006), and Knigge (2006) and color excess used for these distance estimates, and references. All orbital periods are from Ritter and Kolb (2003). Note that for V396 Hya and HV Vir no 2MASS $J$, $H$, $K_S$ magnitudes are available. GP Com is an AM CVn with an orbital period below the lower limit used by Knigge in his semi-empirical donor sequence. SS Cyg, Z Cam, RU Peg, and AE Aqr have orbital periods above the upper limit of the sequence.

a Thorstensen (2003).  b Thorstensen et al. (2008).  c Harrison et al. (1999).
d Harrison et al. (2004).  e Beuermann et al. (2003).  f Beuermann et al. (2004).
g McArthur et al. (1999).  h McArthur et al. (2001).  i Duerbeck (1999).
j Ringwald et al. (1996).  k Bruch and Engel (1994).

| RESIDUALS - % Difference | Ak et al. | Beuermann | Knigge |
|---|---|---|---|
| AVE | –3.7 | 52.0 | –26.4 |
| AVE (< 2hr) | –17.8 | 83.6 | –35.7 |
| AVE (> 3hr) | 11.2 | 18.5 | –13.2 |

Table 2: Table of the percent differences of residuals between trigonometric parallax distances and the near-infrared photometric distance estimates of Ak et al. (2007), Beuermann (2006), and Knigge (2006, 2007). Overall averages, as well as averages below and above the period gap, are given.